\documentclass[preprint,showpacs,preprintnumbers,amsmath,amssymb]{revtex4}
\usepackage{amssymb}
\usepackage{amsmath}
\usepackage{graphicx}
\usepackage{epsfig}
\usepackage{subfigure}
\usepackage{amsfonts}
\usepackage{CJK}

\begin{document}

\title{Universal controlled-phase gate with cat-state qubits in circuit QED}
\author{Yu Zhang$^{1}$, Xiong Zhao$^{1}$, Li Yu$^{1,2}$, Qi-Ping Su$^{1}$}
\author{Chui-Ping Yang$^{1}$}
\email{yangcp@hznu.edu.cn}
\address{$^1$Department of Physics, Hangzhou Normal University, Hangzhou, Zhejiang 310036, China}
\address{$^2$CAS Key Laboratory of Quantum Information, University of Science and Technology of China, Hefei 230026, China}
\date{\today}

\begin{abstract}
 Cat-state qubits (qubits encoded with cat states) have recently drawn intensive attention due to their enhanced life times with quantum error correction. We here propose a method to implement a universal controlled-phase gate of two cat-state qubits, via two microwave resonators coupled to a superconducting transmon qutrit. During the gate operation, the qutrit remains in the ground state; thus decoherence from the qutrit is greatly suppressed. This proposal requires only two basic operations and neither classical pulse nor measurement is needed; therefore the gate realization is simple. Numerical simulations show that high-fidelity implementation of this gate is feasible with current circuit QED technology. The proposal is quite general and can be applied to implement the proposed gate with two microwave resonators or two optical cavities coupled to a single three-level natural or artificial atom.
\end{abstract}

 \maketitle
\date{\today}

\begin{center}
\textbf{I. INTRODUCTION}
\end{center}

Circuit quantum electrodynamics (QED), composed of superconducting (SC)
qubits and microwave resonators or cavities, has developed fast in the past
decade. The circuit QED is considered as one of the most feasible candidates
for quantum information processing (QIP) [1-4]. Due to controllability of
their level spacings, scalability of the circuits, and improvement of
coherence times [5-12], SC qubits are of great importance in QIP. The strong
coupling and ultrastrong coupling between a SC qubit and a microwave
resonator have been demonstrated in experiments [13,14]. In addition, a
coplanar waveguide microwave resonator with a (loaded) quality factor $%
Q=10^{6}$ [15,16] and a three-dimensional microwave resonator with a
(loaded) quality factor $Q\sim3.5 \times 10^{7}$ [17] have been reported in
experiments. A microwave resonator or cavity with a high quality factor can
act as a quantum data bus [18-20] and be used as a quantum memory [21,22],
because it contains microwave photons whose life times are much longer than
that of a SC qubit [23]. Recently, quantum state engineering and QIP with
microwave fields or photons have attracted considerable interest.

Many theoretical proposals have been presented for preparation of Fock
states, squeezed states, coherent states, schr$\ddot{o}$dinger cat states,
and an arbitrary superposition of Fock states of a single microwave
resonator [24-27]. Also, a Fock state and a superposition of Fock states of
a single microwave resonator has been created experimentally [21,28,29] For
two microwave resonators, theoretical proposals have been proposed for
generation of nonclassical microwave field in two resonators [30-33],
construction of two-qubit controlled-phase gates with microwave photons in
two resonators [34], and implementation of quantum state transfer between
microwave photons in two resonators [35-37]. Experimentally, the creation of
N-photon NOON states in two microwave resonators has been reported [38]. A
complete quantum state transfer of a microwave photon qubit between two
resonators can be experimentally realized, by combination of two previous
experiments [39,40] which employed the transfer protocol proposed in
Ref.~[36]. Moreover, schemes have been proposed for generation of
multipartite entangled states of microwave photons in multiple resonators
[41] and creation of entangled coherent states of microwave fields in many
resonators or cavities [42].

The focus of this work is on QIP with cat-state qubits (qubits encoded with
cat states). Cat-state qubits have drawn much attention due to their
enhanced life time with quantum error correction (QEC). For instance, Ofek
\textit{et al}. have made the lifetime of a cat-state qubit up to 320 $\mu$s
with QEC [43]. Recently, there is an increasing interest in QIP with
cat-state encoding qubits. Mirrahimi \textit{et al.} have presented
approaches to realize a set of universal gates on a single cat-state qubit
as well as an entangling gate for creating a Bell state of two cat-state
qubits [44]. Nigg has proposed a method for a deterministic Hadamard gate on
a single cat-state qubit [45]. Heeres \textit{et al}. have experimentally
implemented a set of universal gate on a single cat-state qubit [46]. Yang
\textit{et al}. have proposed a scheme for implementing a SWAP gate of two
cat-state qubits [47]. Moreover, Wang \textit{et al.} have experimentally
generated an entangled Bell state with two cat-state qubits [48]. However,
after a deep search of literature, we found that how to realize a
controlled-phase gate of two \textit{cat-state} qubits has not been
investigated so far. As is well known, a two-qubit controlled phase gate is
\textit{universal}, because two-qubit controlled phase gates, together with
single-qubit gates, form the building blocks of quantum information
processors.

In this paper, we propose a method to realize a universal two-qubit
controlled-phase gate with cat-state qubits, via two microwave resonators
coupled to a SC transmon qutrit (a three-level artificial atom) (Fig.~1).
During the gate operation, the qutrit stays in the ground state; thus
decoherence from the qutrit is greatly suppressed. The gate implementation
is simple because only two basic operations are needed and no classical
pulse or measurement is required. Our numerical simulations show that
high-fidelity implementation of this gate is feasible with current circuit
QED technology.

This paper is organized as follows. In Sec. II, we explicitly show how to
realize a universal controlled-phase gate of two cat-state qubits. In Sec.
III, we numerically calculate the fidelity and briefly discuss the
experimental feasibility. We end up with a conclusion in Sec. IV.

\begin{center}
\textbf{II. CONTROLLED-PHASE GATE OF CAT-STATE QUBITS}
\end{center}

Consider a system consisting of two microwave resonators coupled to a
transmon qutrit (Fig.~1). The three level of the qutrit are labeled as $%
|g\rangle $, $|e\rangle $ and $|f\rangle $, as shown in Fig. 2. It is worth
noting that for an ideal transmon, the $|g\rangle $ $\leftrightarrow $ $%
|f\rangle $ coupling is theoretically zero due to the selection rule [49];
however in practice, there exists a weak coupling between these two states
[50]. Supposed that resonator $a$ is off-resonantly coupled to the $%
|g\rangle \leftrightarrow |e\rangle $ transition of the qutrit with coupling
constant $g$ while resonator $b$ is off-resonantly coupled to the $|e\rangle
\leftrightarrow |f\rangle $ transition of the qutrit with coupling constant $%
\mu$ (Fig.~2). In addition, assume that resonator $a$ is highly detuned
(decoupled) from the $|e\rangle \leftrightarrow |f $ transition of the
qutrit and resonator $(b)$ is highly detuned (decoupled) from the $|g\rangle
\leftrightarrow |e\rangle $ transition of the qutrit (Fig.~3). Note that
these conditions can be achieved by prior adjustment of the level spacings
of the qutrit or/and the resonator frequency. Under these considerations,
the Hamiltonian of the whole system, in the interaction picture and after
making the rotating-wave approximation (RWA), can be written as (in units of
$\hbar =1$)
\begin{equation}
H_{\mathrm{I,1}}=g(e^{i\delta _{a}t}\hat{a}\sigma _{eg}^{+}+h.c.)+\mu
(e^{i\delta _{b}t}\hat{b}\sigma _{fe}^{+}+h.c.),
\end{equation}%
where $\sigma _{eg}^{+}=|e\rangle \langle g|$, $\sigma _{fe}^{+}=|f\rangle
\langle e|$, $\delta _{a}=\omega _{eg}-\omega _{a}<0$ and $\delta
_{b}=\omega _{fe}-\omega _{b}>0.$ The detunigs $\left\vert \delta
_{a}\right\vert $ and $\left\vert \delta _{b}\right\vert $ in Fig. 2 are
given by $\left\vert \delta _{a}\right\vert =\omega _{a}-\omega _{eg}$ and $%
\left\vert \delta _{b}\right\vert =\omega _{fe}-\omega _{b}$. Here, $\hat{a}%
^{+}$ ($\hat{b}^{+}$) is the photon creation operator of resonator $a$ $(b)$%
, $\omega _{fe}$ $(\omega _{eg})$ is the $|e\rangle \leftrightarrow
|f\rangle (|g\rangle \leftrightarrow |e\rangle )$ transition frequency of
the qutrit, while $\omega _{a}$ $(\omega _{b})$ is the frequency of
resonator $a$ $(b)$.

\begin{figure}[tbp]
\begin{center}
\includegraphics[bb=108 606 435 712, width=9.5 cm, clip]{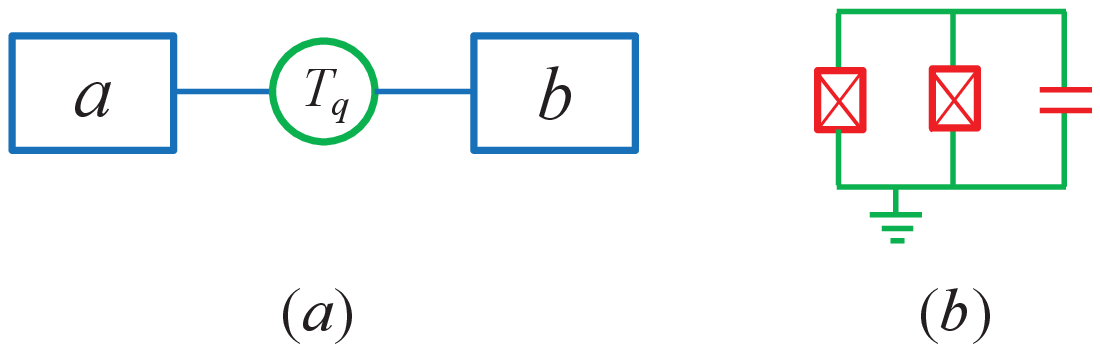} \vspace*{%
-0.08in}
\end{center}
\caption{(Color online) (a) Diagram of two microwave resonators $a$ and $b$
coupled to a transmon qutrit ($T_{q}$). Each resonator can be
one-dimensional or three-dimensional resonator. The qutrit is capacitively
or inductively coupled to each resonator. (b) Electronic circuit of a
transmon qutrit, which consists of two Josephson junctions and a capacitor.}
\label{fig:1}
\end{figure}

\begin{figure}[tbp]
\begin{center}
\includegraphics[bb=181 580 364 716, width=5.0 cm, clip]{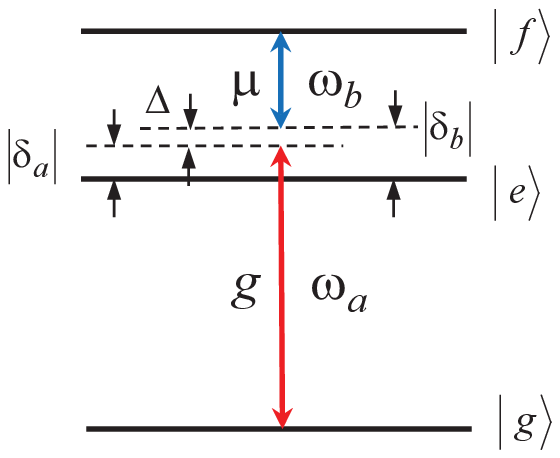} \vspace*{%
-0.08in}
\end{center}
\caption{(Color online) Resonator $a$ is far-off resonant with the $%
|g\rangle\leftrightarrow|e\rangle$ transition of the qutrit with coupling
strength $g$ and detuning $\left\vert \protect\delta _{a}\right\vert $,
while resonator $b$ is far-off resonant with the $|e\rangle\leftrightarrow|f%
\rangle$ transition of the qutrit with coupling strength $\protect\mu$ and
detuning $\left\vert \protect\delta _{b}\right\vert $. Here, $\left\vert
\protect\delta _{a}\right\vert =\protect\omega_{a}-\protect\omega_{eg}$, $%
\left\vert \protect\delta _{b}\right\vert =\protect\omega_{fe}-\protect\omega%
_{b}$, with $\protect\omega _{eg}$ ($\protect\omega _{fe}$) being the $%
|g\rangle \leftrightarrow |e\rangle $ ($|e\rangle \leftrightarrow |f\rangle
) $ transition frequency of the qutrit, while $\protect\omega _{a}$ ($%
\protect\omega _{b}$) being the frequency of resonator $a$ ($b$). In
addition, $\Delta=\left\vert \protect\delta _{b}\right\vert-\left\vert
\protect\delta _{a}\right\vert$. Note that the red vertical line represents
the frequency $\protect\omega_a$ of resonator $a$ while the blue vertical
line represents the frequency $\protect\omega_b$ of resonator $b$. }
\label{fig:2}
\end{figure}

\begin{figure}[tbp]
\begin{center}
\includegraphics[bb=211 426 362 592, width=4.2 cm, clip]{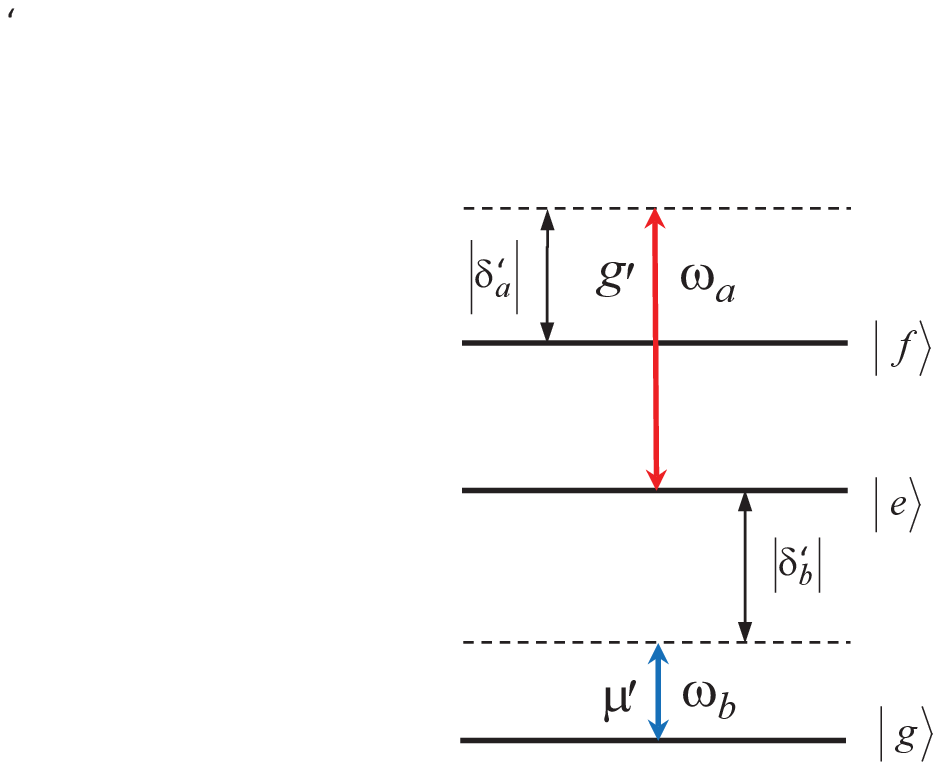} \vspace*{%
-0.08in}
\end{center}
\caption{(Color online) Illustration of resonator $a$ ($b$) is highly
detuned (decoupled) from the $|e\rangle\leftrightarrow|f\rangle$ $%
(|g\rangle\leftrightarrow|e\rangle)$ transition of the qutrit. The high
detuning (or decoupling) can be made by prior adjustment of the level
spacings of the transmon qutrit or/and the frequency of resonator $a$ ($b$),
such that $\left\vert \protect\delta _{a}^{\prime }\right\vert \ggg
g^{\prime }$ and $\left\vert \protect\delta _{b}^{\prime }\right\vert \ggg
\protect\mu ^{\prime }$. Here, $g^{\prime }$ is the coupling constant
between resonator $a$ and the $|e\rangle\leftrightarrow|f\rangle$
transition, $\protect\mu ^{\prime }$ is the coupling constant between
resonator $b$ and the $|g\rangle\leftrightarrow|e\rangle$ transition, $%
\left\vert \protect\delta _{a}^{\prime }\right\vert =\protect\omega_a-%
\protect\omega_{fe}$ is the detuning between the frequency of resonator $a$
and the $|e\rangle\leftrightarrow|f\rangle$ transition frequency, and $%
\left\vert \protect\delta _{b}^{\prime }\right\vert =\protect\omega_{eg}-%
\protect\omega_b$ is the detuning between the frequency of resonator $b$ and
the $|g\rangle\leftrightarrow|e\rangle$ transition frequency. Note that the
coupling of both resonators with the $|g\rangle\leftrightarrow|f\rangle$
transition of the qutrit is negligible because of the forbidden or very weak
$|g\rangle\leftrightarrow|f\rangle$ transition [49,50].}
\label{fig:3}
\end{figure}

Under the large-detuning conditions $\left\vert \delta _{a}\right\vert \gg g$
and $\left\vert \delta _{b}\right\vert \gg \mu $, the Hamiltonian (1)
becomes [46]
\begin{align}
H_{\mathrm{e}}=& -\lambda _{a}(\hat{a}^{+}\hat{a}|g\rangle \langle g|-\hat{a}%
\hat{a}^{+}|e\rangle \langle e|)  \notag \\
& -\lambda _{b}(\hat{b}^{+}\hat{b}|e\rangle \langle e|-\hat{b}\hat{b}%
^{+}|f\rangle \langle f|)  \notag \\
& +\lambda (e^{-i\bigtriangleup t}\hat{a}^{+}\hat{b}^{+}\sigma
_{fg}^{-}+h.c.),
\end{align}%
where $\lambda _{a}=g^{2}/\delta _{a}$, $\lambda _{b}=\mu ^{2}/\delta _{b}$,
$\lambda =\left( g\mu /2\right) (1/|\delta _{a}|+1/|\delta _{b}|)$, $%
\bigtriangleup =\left\vert \delta _{b}\right\vert -|\delta _{a}|$, and $%
\sigma _{fg}^{-}=|g\rangle \langle f|$. The first four terms of Eq.(2)
describe the photon-number dependent stark shifts of the energy levels $%
|g\rangle $, $|e\rangle $ and $|f\rangle $, while the last two terms
describe the $|f\rangle $ $\leftrightarrow $ $|g\rangle $ coupling caused
due to the two-resonator cooperation. For $|\bigtriangleup |\gg \{\lambda
_{a},\lambda _{b},\lambda \}$, the effective Hamiltonian $H_{\mathrm{e}}$
changes to [51]
\begin{align}
H_{\mathrm{e}}=& -\lambda _{a}(\hat{a}^{+}\hat{a}|g\rangle \langle g|-\hat{a}%
\hat{a}^{+}|e\rangle \langle e|)  \notag \\
& -\lambda _{b}(\hat{b}^{+}\hat{b}|e\rangle \langle e|-\hat{b}\hat{b}%
^{+}|f\rangle \langle f|)  \notag \\
& +\chi (\hat{a}\hat{a}^{+}\hat{b}\hat{b}^{+}|f\rangle \langle f|-\hat{a}^{+}%
\hat{a}\hat{b}^{+}\hat{b}|g\rangle \langle g|),
\end{align}%
where $\chi =\lambda ^{2}/\Delta $. From Eq.~(3) one can see that each term
is associated with the level $|g\rangle $, $|e\rangle $, or $|f\rangle $.
When the levels $|e\rangle $ and $|f\rangle $ are not occupied, they will
remain unpopulated under the Hamiltonian (3). In this case, the effective
Hamiltonian (3) reduces to
\begin{equation}
H_{\mathrm{e}}=H_{0}+H_{\mathrm{int}},
\end{equation}%
with
\begin{eqnarray}
H_{0} &=&-\lambda _{a}\hat{a}^{+}\hat{a}|g\rangle \langle g|=-\lambda _{a}%
\hat{n}_{a}|g\rangle \langle g|,  \notag \\
H_{\mathrm{int}} &=&-\chi \hat{a}^{+}\hat{a}\hat{b}^{+}\hat{b}|g\rangle
\langle g|=-\chi \hat{n}_{a}\hat{n}_{b}|g\rangle \langle g|,
\end{eqnarray}%
where $\hat{n}_{a}=\hat{a}^{+}\hat{a}$ $(\hat{n}_{b}=\hat{b}^{+}\hat{b})$ is
the photon number operator for resonator $a$ $(b)$. Because of $[H_{0},H_{%
\mathrm{int}}]=0$, the unitary operator $U_{1}=e^{-iH_{e}t}$ can be written
as
\begin{equation}
U_{1}=e^{-iH_{0}t}e^{-iH_{\mathrm{int}}t}=\exp \left( i\lambda _{a}\hat{n}%
_{a}|g\rangle \langle g|t\right) \exp \left( i\chi \hat{n}_{a}\hat{n}%
_{b}|g\rangle \langle g|t\right) .
\end{equation}

The two logical states $|0\rangle $ and $|1\rangle $ of a cat-state qubit
are encoded with cat states of a resonator, i.e., $|0\rangle =M_{\alpha
}^{+}(|\alpha \rangle +|-\alpha \rangle )$ and $|1\rangle =M_{\alpha
}^{-}(|\alpha \rangle -|-\alpha \rangle )$, respectively. Here, $M_{\alpha
}^{\pm }=1/\sqrt{2(1\pm e^{-2|\alpha |^{2}})}$ are normalization
coefficients. In terms of $|\alpha \rangle =e^{-|\alpha
|^{2}/2}\sum\limits_{n=0}^{\infty }\frac{\alpha ^{n}}{\sqrt{n!}}|n\rangle $
and $|-\alpha \rangle =e^{-|\alpha |^{2}/2}\sum\limits_{n=0}^{\infty }\frac{%
(-\alpha )^{n}}{\sqrt{n!}}|n\rangle $, we have
\begin{equation}
|0\rangle =\sum\limits_{m=0}^{\infty }C_{2m}|2m\rangle ,\ \ |1\rangle
=\sum\limits_{n=0}^{\infty }C_{2n+1}|2n+1\rangle ,
\end{equation}%
where $C_{2m}=2M_{\alpha }^{+}e^{-|\alpha |^{2}/2}\alpha ^{2m}/\sqrt{(2m)!}$
and $C_{2n+1}=2M_{\alpha }^{-}e^{-|\alpha |^{2}/2}\alpha ^{2n+1}/\sqrt{%
(2n+1)!}$. From Eq.~(7), one can see that the state $|0\rangle $ is
orthogonal to the state $|1\rangle $, which is independent of $\alpha $
(except for $\alpha =0$).

The four logical states of two cat-state qubits are $|00\rangle _{ab}$, $%
|01\rangle _{ab}$, $|10\rangle _{ab}$ and $|11\rangle _{ab}$, where the left
0 and 1 are encoded with cat states of resonator $a$ while the right 0 and 1
are encoded with cat states of resonator $b$. Suppose that the qutrit is
initially in the ground state $|g\rangle $. For an interaction time $t=t_{1}$%
, the unitary operation $U_{1}$ leads to the following state transformations
(see Appendix for details)
\begin{align}
U_{1}|00\rangle _{ab}|g\rangle & =\sum\limits_{m,m^{\prime }=0}^{\infty
}F_{1}(m,m^{\prime },t_{1})C_{2m}C_{2m^{\prime }}|2m\rangle _{a}|2m^{\prime
}\rangle _{b}|g\rangle ,  \notag \\
U_{1}|01\rangle _{ab}|g\rangle & =\sum\limits_{m,n^{\prime }=0}^{\infty
}F_{2}(m,n^{\prime },t_{1})C_{2m}C_{2n^{\prime }+1}|2m\rangle
_{a}|2n^{\prime }+1\rangle _{b}|g\rangle ,  \notag \\
U_{1}|10\rangle _{ab}|g\rangle & =\sum\limits_{n,m\prime =0}^{\infty
}F_{3}(n,m^{\prime },t_{1})C_{2n+1}C_{2m^{\prime }}|2n+1\rangle
_{a}|2m^{\prime }\rangle _{b}|g\rangle ,  \notag \\
U_{1}|11\rangle _{ab}|g\rangle & =\sum\limits_{n,n^{\prime }=0}^{\infty
}F_{4}(n,n^{\prime },t_{1})C_{2n+1}C_{2n^{\prime }+1}|2n+1\rangle
_{a}|2n^{\prime }+1\rangle _{b}|g\rangle ,
\end{align}%
with%
\begin{eqnarray}
F_{1}(m,m^{\prime },t_{1}) &=&\exp (i\lambda _{a}2mt_{1})\exp \left[
i(2m)(2m^{\prime })\chi t_{1}\right] ,  \notag \\
F_{2}(m,n^{\prime },t_{1}) &=&\exp (i\lambda _{a}2mt_{1})\exp
[i(2m)(2n^{\prime }+1)\chi t_{1}],  \notag \\
F_{3}(n,m^{\prime },t_{1}) &=&\exp [i\lambda _{a}(2n+1)t_{1}]\exp
[i(2n+1)(2m^{\prime })\chi t_{1}],  \notag \\
F_{4}(n,n^{\prime },t_{1}) &=&\exp [i\lambda _{a}(2n+1)t_{1}]\exp
[i(2n+1)(2n^{\prime }+1)\chi t_{1}].
\end{eqnarray}

We now adjust the frequency of resonator $a$ such that resonator $a$ is
far-off resonant with the $|g\rangle \leftrightarrow |e\rangle $ transition
of the qutrit with coupling strength $\widetilde{g}$ and detuning $|%
\widetilde{\delta }_{a}|$ (Fig. 4), while it is highly detuned (decoupled)
from the $|e\rangle \leftrightarrow |f\rangle $ transition (Fig. 5). Here, $|%
\widetilde{\delta }_{a}|=\omega _{eg}-\tilde{\omega}_{a}$ (Fig. 4), with $%
\widetilde{\omega }_{a}$ being the adjusted frequency of resonator $a$. In
addition, adjust the frequency of resonator $b$ such that resonator $b$ is
decoupled from the qutrit. Note that the frequency of a microwave resonator
can be rapidly adjusted with a few nanoseconds [52,53]. Under these
considerations, the Hamiltonian in the interaction picture and after making
the RWA is given by

\begin{figure}[tbp]
\begin{center}
\includegraphics[bb=192 581 361 713, width=5.0 cm, clip]{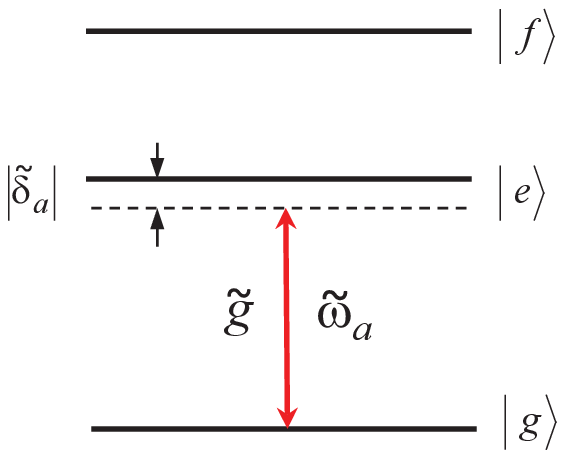} \vspace*{%
-0.08in}
\end{center}
\caption{(Color online) Resonator $a$ is far-off resonant with the $%
|g\rangle\leftrightarrow|e\rangle$ transition of the qutrit with coupling
strength $\widetilde{g}$ and detuning $\left\vert \widetilde{\protect\delta}%
_{a}\right\vert $. Here, $\left\vert \widetilde{\protect\delta}%
_{a}\right\vert=\protect\omega_{eg}-\widetilde{\protect\omega}_{a}$, with $%
\widetilde{\protect\omega}_{a}$ being the adjusted frequency of resonator $a$
(labelled by the vertical line). The frequency of resonator $b$ is adjusted
such that resonator $b$ is decoupled from the qutrit. Note that the
dispersive qutrit-cavity coupling with a detuning $\left\vert \widetilde{%
\protect\delta}_{a}\right\vert $ illustrated here can also be obtained by
adjusting the level spacings of the qutrit but the cavity frequency being
fixed.}
\label{fig:4}
\end{figure}

\begin{figure}[tbp]
\begin{center}
\includegraphics[bb=192 578 370 737, width=5.0 cm, clip]{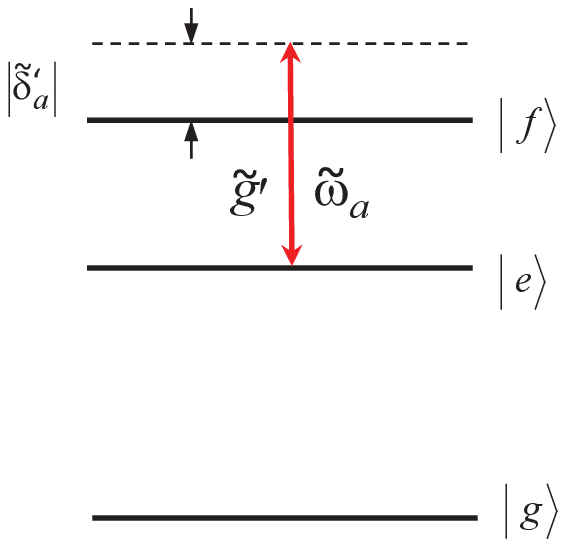} \vspace*{%
-0.08in}
\end{center}
\caption{(Color online) Illustration of resonator $a$ is highly detuned
(decoupled) from the $|e\rangle\leftrightarrow|f\rangle$ transition of the
qutrit. The decoupling can be made as long as the condition $\left\vert
\widetilde{\protect\delta }_{a}^{\prime }\right\vert \ggg \widetilde{g}%
^{\prime }$ can be satisfied. Here, $\widetilde{g}^{\prime }$ is the
coupling constant between resonator $a$ and the $|e\rangle\leftrightarrow|f%
\rangle$ transition, while $\left\vert \widetilde{\protect\delta }%
_{a}^{\prime }\right\vert =\widetilde{\protect\omega }_{a}-\protect\omega %
_{fe}$ is the detuning between the frequency of resonator $a$ and the $%
|e\rangle\leftrightarrow|f\rangle$ transition frequency. Note that the
coupling of resonator $a$ with the $|g\rangle\leftrightarrow|f\rangle$
transition of the qutrit is negligible because of the forbidden or very weak
$|g\rangle\leftrightarrow|f\rangle$ transition [49,50]. In addition, since
the frequency of resonator $b$ is far detuned, resonator $b$ is decoupled
from the qutrit.}
\label{fig:5}
\end{figure}

\begin{equation}
H_{\mathrm{I,2}}=\tilde{g}(e^{i\widetilde{\delta }_{a}t}\hat{a}\sigma
_{eg}^{+}+h.c.),
\end{equation}%
where $\tilde{\delta}_{a}=|\widetilde{\delta _{a}}|=\omega _{eg}-\tilde{%
\omega}_{a}>0.$

For $\tilde{\delta}_{a}\gg \tilde{g}$ and the level $|e\rangle $ being not
occupied, we have
\begin{equation}
\tilde{H}_{\mathrm{e}}=-\widetilde{\lambda }_{a}\hat{n}_{a}|g\rangle \langle
g|,
\end{equation}%
with $\widetilde{\lambda }_{a}=\tilde{g}^{2}/\tilde{\delta}_{a}$. Then,
performing a unitary transformation $U_{2}=\exp (i\widetilde{\lambda }_{a}%
\hat{n}_{a}|g\rangle \langle g|t_{2})$ for an interaction time $t=t_{2}$, we
obtain from Eqs.~(8) and (9)
\begin{align}
U_{2}U_{1}|00\rangle _{ab}|g\rangle & =\sum\limits_{m,m^{\prime }=0}^{\infty
}\widetilde{F}_{1}(m,m^{\prime },t_{1})C_{2m}C_{2m^{\prime }}|2m\rangle
_{a}|2m^{\prime }\rangle _{b}|g\rangle ,  \notag \\
U_{2}U_{1}|01\rangle _{ab}|g\rangle & =\sum\limits_{m,n^{\prime }=0}^{\infty
}\widetilde{F}_{2}(m,n^{\prime },t_{1})C_{2m}C_{2n^{\prime }+1}|2m\rangle
_{a}|2n^{\prime }+1\rangle _{b}|g\rangle ,  \notag \\
U_{2}U_{1}|10\rangle _{ab}|g\rangle & =\sum\limits_{n,m^{\prime }=0}^{\infty
}\widetilde{F}_{3}(n,m^{\prime },t_{1})C_{2n+1}C_{2m^{\prime }}|2n+1\rangle
_{a}|2m^{\prime }\rangle _{b}|g\rangle ,  \notag \\
U_{2}U_{1}|11\rangle _{ab}|g\rangle & =\sum\limits_{n,n^{\prime }=0}^{\infty
}\widetilde{F}_{4}(n,n^{\prime },t_{1})C_{2n+1}C_{2n^{\prime
}+1}|2n+1\rangle _{a}|2n^{\prime }+1\rangle _{b}|g\rangle .
\end{align}%
with%
\begin{eqnarray}
\widetilde{F}_{1}(m,m^{\prime },t_{1}) &=&\exp [i2m(\lambda _{a}t_{1}+%
\widetilde{\lambda }_{a}t_{2})]\exp [i\left( 2m\right) \left( 2m^{\prime
}\right) \chi t_{1}],  \notag \\
\widetilde{F}_{2}(m,n^{\prime },t_{1}) &=&\exp [i2m(\lambda _{a}t_{1}+%
\widetilde{\lambda }_{a}t_{2})]\exp [i(2m)(2n+1)\chi t_{1}],  \notag \\
\widetilde{F}_{3}(n,m^{\prime },t_{1}) &=&\exp [i(2n+1)(\lambda _{a}t_{1}+%
\widetilde{\lambda }_{a}t_{2})]\exp [i(2n+1)(2m^{\prime })\chi t_{1}],
\notag \\
\widetilde{F}_{4}(n,n^{\prime },t_{1}) &=&\exp [i(2n+1)(\lambda _{a}t_{1}+%
\widetilde{\lambda }_{a}t_{2})]\exp [i(2n+1)(2n^{\prime }+1)\chi t_{1}].
\end{eqnarray}

Note that the index factors $\left( 2m\right) \left( 2m^{\prime }\right) $, $%
(2m)(2n+1)$, and $(2n+1)(2m^{\prime })$ of Eq.~(13) are even numbers, while
the index factor $(2n+1)(2n^{\prime }+1)$ is an odd number. By setting $%
\lambda _{a}=-\widetilde{\lambda }_{a}$ (i.e., $g^{2}/\delta _{a}=-\tilde{g}%
^{2}/\tilde{\delta}_{a}$) and $t_{2}=t_{1}=\pi /\left\vert \chi \right\vert $%
, we have $\widetilde{F}_{1}(m,m^{\prime },t_{1})=\widetilde{F}%
_{2}(m,n^{\prime },t_{1})=\widetilde{F}_{3}(n,m^{\prime },t_{1})=1$ but $%
\widetilde{F}_{4}(n,n^{\prime },t_{1})=-1.$ Hence, the states (12) become
\begin{align}
U_{2}U_{1}|00\rangle _{ab}|g\rangle & =|00\rangle _{ab}|g\rangle ,  \notag \\
U_{2}U_{1}|01\rangle _{ab}|g\rangle & =|01\rangle _{ab}|g\rangle ,  \notag \\
U_{2}U_{1}|10\rangle _{ab}|g\rangle & =|10\rangle _{ab}|g\rangle ,  \notag \\
U_{2}U_{1}|11\rangle _{ab}|g\rangle & =-|11\rangle _{ab}|g\rangle ,
\end{align}%
which shows that the above two basic operations (i.e., $U_{1}$ and $U_{2}$)
have completed a universal controlled-phase gate of two cat-state qubits,
described by $|00\rangle _{ab}\rightarrow |00\rangle _{ab}$, $|01\rangle
_{ab}\rightarrow |01\rangle _{ab}$, $|10\rangle _{ab}\rightarrow |10\rangle
_{ab}$, and $|11\rangle _{ab}\rightarrow -|11\rangle _{ab}$. After this
gate, an arbitrary pure state of two cat-state qubits, given by $|\phi
\rangle _{ab}=\alpha |00\rangle _{ab}+\beta |01\rangle _{ab}+\gamma
|10\rangle _{ab}+\zeta |11\rangle _{ab}$, is transformed as follows
\begin{equation}
|\phi \rangle _{ab}\rightarrow \alpha |00\rangle _{ab}+\beta |01\rangle
_{ab}+\gamma |10\rangle _{ab}-\zeta |11\rangle _{ab}.
\end{equation}

From description given above, one can see that the qutrit remains in the
ground state during the entire operation. Hence, decoherence from the qutrit
is greatly suppressed.

As shown above, the Hamiltonian (11) for the second unitary operation ($%
U_{2} $) was constructed by tuning cavity frequency. However, we point out
that tuning cavity frequency is unnecessary. Alternatively, one can obtain
the Hamiltonian (11) by adjusting the level spacings of the qutrit to meet
the conditions required for constructing this Hamiltonian (11). Note that
for a SC qutrit, the level spacings can be rapidly (within 1-3 ns) adjusted
by varying external control parameters (e.g., magnetic flux applied to the
superconducting loop of a SC phase, transmon [54], Xmon [10], or flux
qubit/qutrit [55]).

We should mention that the Hamiltonian (4) was previously proposed to
realize a controlled-phase gate of two \textit{discrete-variable} qubits
[56], for which the two logic states of a qubit are encoded with \textit{the
vacuum state and a single-photon state} of a cavity mode. In stark contrast,
the present work aims at implementing a controlled-phase gate of two \textit{%
continuous-variable} qubits, for which the two logic states of a qubit are
encoded with \textit{cat states} of a resonator or cavity.

\begin{center}
\textbf{III. POSSIBLE EXPERIMENTAL IMPLEMENTATION}
\end{center}

In above, we have explicitly shown how to realize a controlled-phase gate of
two cat-state qubits. We now give a brief discussion on the experimental
feasibility by considering a setup of a SC transmon qutrit coupled to two 3D
microwave resonators or cavities.

From the description given above, one can see that the gate implementation
involves the following two basic operations:

(i) The first operation is described by the Hamiltonian (1). In reality, the
inter-resonator crosstalk between the two resonators is inevitable [57], and
there exist the unwanted coupling of resonator $a$ with the $|e\rangle
\leftrightarrow |f\rangle $ transition and the unwanted coupling of
resonator $b$ with the $|g\rangle \leftrightarrow |e\rangle $ transition of
the qutrit. When these factors are taken into account, the Hamiltonian (1)
becomes
\begin{eqnarray}
\widetilde{H}_{\mathrm{I,}1} &=&g(e^{i\delta _{a}t}\hat{a}\sigma
_{eg}^{+}+h.c.)+\mu (e^{i\delta _{b}t}\hat{b}\sigma _{fe}^{+}+h.c.)  \notag
\\
&&+g^{\prime }(e^{i\delta _{a}^{\prime }t}\hat{a}\sigma _{fe}^{+}+h.c.)+\mu
^{\prime }(e^{i\delta _{b}^{\prime }t}\hat{b}\sigma _{eg}^{+}+h.c.)  \notag
\\
&&+g_{ab}(e^{-i\bigtriangleup _{ab}t}\hat{a}\hat{b}^{+}+h.c.),
\end{eqnarray}%
where the first bracket term represents the interaction of resonator $a$
with the $|g\rangle \leftrightarrow |e\rangle $ transition, the second
bracket term represents the interaction of resonator $b$ with the $|e\rangle
\leftrightarrow |f\rangle $ transition, the third bracket term  represents
the unwanted coupling between resonator $a$ and the $|e\rangle
\leftrightarrow |f\rangle $ transition with coupling strength $g^{\prime }$
and detuning $\delta _{a}^{\prime }=\omega _{fe}-\omega _{a}<0$ (Fig. 3),
and the fourth bracket term represents the unwanted coupling between
resonator $b$ and the $|g\rangle \leftrightarrow |e\rangle $ transition of
the qutrit with coupling strength $\mu ^{\prime }$ and detuning $\delta
_{b}^{\prime }=\omega _{eg}-\omega _{b}>0$ (Fig. 3). In addition, the last
bracket term of Eq. (16) represents the inter-resonator crosstalk, where $%
g_{ab}$ is the coupling strength between the two resonators while $%
\bigtriangleup _{ab}=\omega _{a}-\omega _{b}$ is the difference between the
two-resonator frequencies.

(ii) The second operation is described by the Hamiltonian (10). In practice,
the inter-resonator crosstalk between the two resonators and the unwanted
coupling of resonator $a$ with the $|e\rangle \leftrightarrow |f\rangle $
transition should be considered. Note that for the second operation, the
frequency of resonator $b$ was far detuned such that resonator $b$ is
decoupled from the qutrit. When these factors are taken into account, the
Hamiltonian (10) becomes

\begin{eqnarray}
\widetilde{H}_{\mathrm{I,}2} &=&\tilde{g}(e^{i\widetilde{\delta }_{a}t}\hat{a%
}\sigma _{eg}^{+}+h.c.)+\tilde{g}^{\prime }(e^{i\widetilde{\delta }%
_{a}^{\prime }t}\hat{a}\sigma _{fe}^{+}+h.c.)  \notag \\
&&+\widetilde{g}_{ab}(e^{-i\widetilde{\bigtriangleup }_{ab}t}\hat{a}\hat{b}%
^{+}+h.c.),
\end{eqnarray}%
where the first bracket term represents the interaction of resonator $a$
with the $|g\rangle \leftrightarrow |e\rangle $ transition, while the second
bracket term represents the unwanted coupling between resonator $a$ and the $%
|e\rangle \leftrightarrow |f\rangle $ transition with coupling strength $%
\widetilde{g}^{\prime }$ and detuning $\widetilde{\delta }_{a}^{\prime
}=\omega _{fe}-\widetilde{\omega }_{a}<0$ (Fig. 5). The last bracket term of
Eq. (17) represents the inter-resonator crosstalk, where $\widetilde{g}_{ab}$
is the coupling strength between the two resonators while $\widetilde{%
\bigtriangleup }_{ab}=\widetilde{\omega }_{a}-\widetilde{\omega }_{b}$ is
the difference between the two-resonator frequencies.

The dynamics of the lossy system is determined by
\begin{align}
\frac{d\rho }{dt}=& -i[\widetilde{H}_{\mathrm{I,}i},\rho ]+\kappa _{a}%
\mathcal{L}[a]+\kappa _{b}\mathcal{L}[b]  \notag \\
& +\gamma _{eg}\mathcal{L}[\sigma _{eg}^{-}]+\gamma _{fe}\mathcal{L}[\sigma
_{fe}^{-}]+\gamma _{fg}\mathcal{L}[\sigma _{fg}^{-}]  \notag \\
& +\sum\limits_{j=e,f}\{\gamma _{\varphi j}(\sigma _{jj}\rho \sigma
_{jj}-\sigma _{jj}\rho /2-\rho \sigma _{jj}/2)\},
\end{align}%
where $\widetilde{H}_{\mathrm{I,}i}$ is the full Hamiltonian given above ($%
i=1,2$), $\sigma _{eg}^{-}=|g\rangle \langle e|$, $\sigma
_{fe}^{-}=|e\rangle \langle f|$, $\sigma _{fg}^{-}=|g\rangle \langle f|$, $%
\sigma _{jj}=|j\rangle \langle j|(j=e,f)$; and $\mathcal{L}[\xi ]=\xi \rho
\xi ^{\dag }-\xi ^{\dag }\xi \rho /2-\rho \xi ^{\dag }\xi /2$, with $\xi
=a,b,\sigma _{eg}^{-},\sigma _{fe}^{-},\sigma _{fg}^{-}$. Here, $\kappa
_{a}(\kappa _{b})$ is the photon decay rate of resonator $a$ $(b)$. In
addition, $\gamma _{eg}$ is the energy relaxation rate for the level $%
|e\rangle $ of the qutrit, $\gamma _{fe}(\gamma _{fg})$ is the energy
relaxation rate of the level $|f\rangle $ of the qutrit for the decay path $%
|f\rangle \longrightarrow |e\rangle (|g\rangle )$, and $\gamma _{\varphi j}$
is the dephasing rate of the level $|j\rangle (j=e,f)$ of the qutrit.
\newline

The fidelity of the operations is given by
\begin{equation}
\mathcal{F}=\sqrt{\langle \psi _{\mathrm{id}}|\rho |\psi _{\mathrm{id}%
}\rangle },
\end{equation}%
where $|\psi _{\mathrm{id}}\rangle $ is the output state of an ideal system
without dissipation, dephasing and crosstalk \textit{etc}.; while $\rho $ is
the final practical density operator of the system when the operation is
performed in a realistic situation. For simplicity, choose $\alpha =\cos
\theta \cos \varphi ,$ $\beta =\cos \theta \sin \varphi ,$ $\gamma =\sin
\theta \cos \varphi ,$ and $\zeta =\sin \theta \sin \varphi ,$ which satisfy
the normalization condition $\left\vert \alpha \right\vert ^{2}+\left\vert
\beta \right\vert ^{2}+\left\vert \gamma \right\vert ^{2}+\left\vert \zeta
\right\vert ^{2}=1.$ The initial state of the qutrit-resonator system is
thus written as $|\psi _{\mathrm{in}}\rangle =\left( \cos \theta \cos
\varphi |00\rangle _{ab}+\cos \theta \sin \varphi |01\rangle _{ab}+\sin
\theta \cos \varphi |10\rangle _{ab}+\sin \theta \sin \varphi |11\rangle
_{ab}\right) |g\rangle $. The output state is $|\psi _{\mathrm{id}}\rangle
=\left( \cos \theta \cos \varphi |00\rangle _{ab}+\cos \theta \sin \varphi
|01\rangle _{ab}+\sin \theta \cos \varphi |10\rangle _{ab}-\sin \theta \sin
\varphi |11\rangle _{ab}\right) |g\rangle $. In the following, we will
consider the cases: (i) $\theta =\varphi =\pi /4;$ (ii) $\theta =\varphi
=\pi /3;$ (iii) $\theta =\pi /4,\varphi =\pi /3;$ and (iv) $\theta =\pi
/3,\varphi =\pi /4;$ which correspond to four initial states.

\begin{figure}[tbp]
\begin{center}
\includegraphics[bb=79 54 750 523, width=12.5 cm, clip]{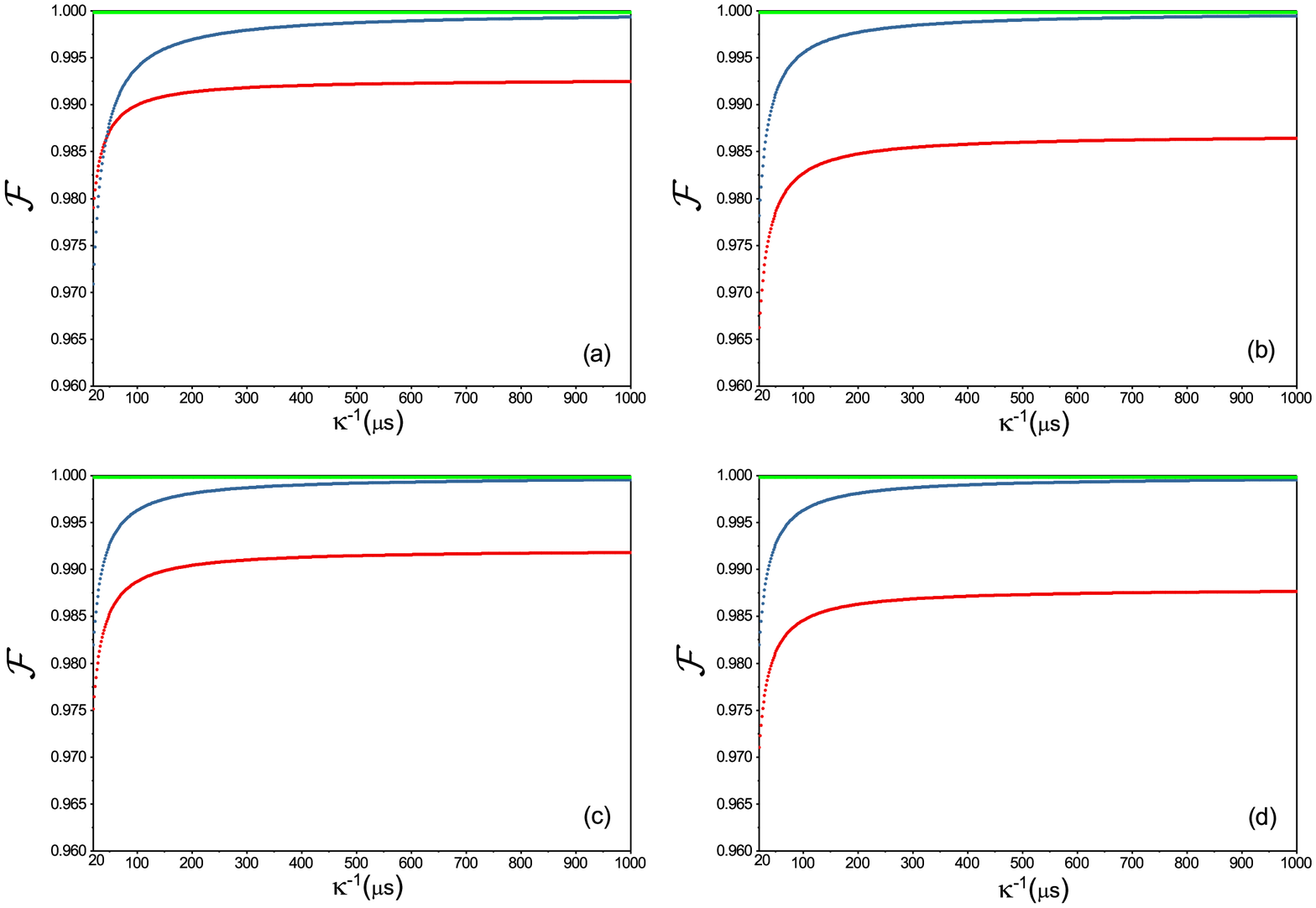} \vspace*{%
-0.08in}
\end{center}
\caption{(Color online) Fidelity versus $\protect\kappa^{-1}$. The plots are
drawn for $\protect\alpha=0.5$. Other parameters used in the numerical
simulation are referred to the text. Green curves are based on the effective
Hamiltonians (4) and (11) but not considering decoherence and the inter-resonator
crosstalk; blue curves are based on the effective Hamitonians (4) and (11) and considering
decoherence and the inter-resonator crosstalk; while red curves
are based on the full Hamiltonians (16) and (17) and taking decoherence and the
inter-resonator crosstalk into account. (a) is plotted for $\protect\theta =\protect%
\varphi =\protect\pi /4$; (b) is for $\protect\theta =\protect\varphi =%
\protect\pi /3$; (c) is for $\protect\theta =\protect\pi /4,$ $\protect%
\varphi =\protect\pi /3$; while (d) is for $\protect\theta =\protect\pi /3,$ $%
\protect\varphi =\protect\pi /4$.}
\label{fig:6}
\end{figure}

For a transmon qutrit, the typical transition frequency between two
neighboring levels can be varied from 3 to 10 GHz. In addition, the
anharmonicity of the level spacings for a transmon qutrit can be made to be
within $100\sim 500$ MHz [14]. As an example, we thus consider $\omega
_{eg}/2\pi =6.5$ GHz and $\omega _{fe}/2\pi =6$ GHz. By choosing $\delta
_{a}/2\pi =-1.0$ GHz and $\delta _{b}/2\pi =1.1$ GHz, we have $\omega
_{a}/2\pi =7.5$ GHz and $\omega _{b}/2\pi =4.9$ GHz, for which we have $%
\triangle _{ab}/2\pi =2.6$ GHz. We set $\widetilde{\delta }_{a}/2\pi =1.0$
GHz, for which we have $\widetilde{\omega }_{a}/2\pi =5.5$ GHz. By choosing $%
\widetilde{\omega }_{b}/2\pi =3.5$ GHz, we have $\widetilde{\triangle }%
_{ab}/2\pi =2$ GHz. In addition, we have $\delta _{a}^{\prime }/2\pi =-1.5$
GHz, $\delta _{b}^{\prime }/2\pi =1.6$ GHz, and $\widetilde{\delta }%
_{a}^{\prime }/2\pi =-0.5$ GHz. Other parameters used in the numerical
simulation are: (i) $\gamma _{eg}^{-1}=60$ $\mu $s, $\gamma _{fg}^{-1}=150$ $%
\mu $s [58], $\gamma _{fe}^{-1}=30$ $\mu $s, $\gamma _{\phi e}^{-1}=\gamma
_{\phi f}^{-1}=20$ $\mu $s, (ii) $g/2\pi =\mu /2\pi =95$ MHz (available in
experiments [14]), and (iii) $\alpha =0.5$. Here, we consider a rather
conservative case for decoherence time of transmon qutrits because energy
relaxation time with a range from 65 $\mu $s to 0.1 ms and dephasing time
from 25 $\mu $s to 70 $\mu $s have been experimentally reported for a 3D
superconducting transmon device [7,11,48]. The value of $\widetilde{g}$ is
determined according to $g^{2}/\delta _{a}=-\tilde{g}^{2}/\tilde{\delta}_{a}$%
, given $g$, $\delta _{a}$, and $\widetilde{\delta }_{a}$. For a transmon
qutrit [49], one has $g^{\prime }\sim \sqrt{2}g,$ $\mu ^{\prime }\sim \mu /%
\sqrt{2},$ and $\widetilde{g}^{\prime }\sim \sqrt{2}\widetilde{g}.$ We set $%
g_{ab}=0.01g$, which can be readily achieved in experiments [33].

For simplicity, assume $\kappa _{a}=\kappa _{b}=\kappa .$ By solving the
master equation (18), we numerically calculate the fidelity versus $\kappa
^{-1}$, as shown in Fig.~6. Fig. 6(a) is plotted for $\theta =\varphi =\pi
/4.$ Fig. 6(b) is for $\theta =\varphi =\pi /3.$ Fig. 6(c) is for $\theta
=\pi /4,\varphi =\pi /3.$ Fig. 6(d) is for $\theta =\pi /3,\varphi =\pi /4$.
The red curves in Fig. 6 are drawn by numerical simulations, which are based
on the full Hamiltonians $\widetilde{H}_{\mathrm{I,}1}$ in Eq. (16)\ and $%
\widetilde{H}_{\mathrm{I,}2}$ in Eq. (17) and take decoherence and the inter-resonator crosstalk into account.
The red curves illustrate that when $\kappa ^{-1}\geq 300$ $\mu $, fidelity
exceeds: (i) 0.9918 for $\theta =\varphi =\pi /4;$ (ii) 0.9854 for $\theta
=\varphi =\pi /3;$ (iii) 0.9910 for $\theta =\pi /4,\varphi =\pi /3;$ and
(iv) 0.9868 for $\theta =\pi /3,\varphi =\pi /4.$ These results imply that
the fidelity depends on the choice of the initial state of the two
resonators and a high fidelity can be obtained when the gate is  
performed in a realistic situation. 

To see how good the approximations are, we have calculated the fidelity
based on the effective Hamiltonians given in Eq. (4) and Eq. (11) and
by considering decoherence and the inter-resonator crosstalk (see the blue
curves in Fig. 6). From the red curves and the bule curves depicted in Fig.
6, one can see that compared to the case of the gate being performed based
on the effective Hamiltonians, the fidelity for the gate performed in a
realistic situation is slightly decreased by $0.9\%-1.5\%.$ This implies
that the approximations made for the effective Hamiltonians are reasonable.

Lifetime $\sim 1$ ms of microwave photons has been experimentally
demonstrated in a coaxial resonator [17,48]. For $\kappa ^{-1}=300$ $\mu $s,
we have $Q_{a}=1.2\times 10^{7}$ for $\omega _{a}/2\pi =6.5$ GHz, $%
\widetilde{Q}_{a}=1.0\times 10^{7}$ for $\widetilde{\omega }_{a}/2\pi =5.5$
GHz, $Q_{b}=9.2\times 10^{6}$ for $\omega _{b}/2\pi =4.9$ GHz, and $%
\widetilde{Q}_{b}=6.6\times 10^{6}$ for $\widetilde{\omega }_{b}/2\pi =3.5$
GHz. Note that a high quality factor $Q=3.5\times 10^{7}$ of a 3D
superconducting resonator has been experimentally demonstrated [17]. The
analysis here implies that the high-fidelity implementation of the proposed
gate is feasible within the current circuit QED technology.

\begin{center}
\textbf{IV. CONCLUSIONS}
\end{center}

We have proposed a method to realize a universal controlled-phase gate of
two cat-state qubits, via two microwave resonators coupled to a
superconducting transmon qutrit. This method can be extended to a wide range
of physical systems such as two microwave or optical cavities coupled to a
single three-level natural or artificial atom. As shown above, this proposal
has these features. During the gate operation, the qutrit remains in the
ground state; thus decoherence from the qutrit is greatly suppressed.
Because only two basic operations are needed and neither classical pulse nor
measurement is required, the gate realization is simple. Our numerical
simulations show that high-fidelity implementation of the proposed gate is
feasible with current circuit QED technology. To the best of our knowledge,
this work is the first to demonstrate the implementation of a
controlled-phase gate with cat-state qubits based on cavity- or circuit-QED.
We hope that this work will stimulate experimental activities in the near
future.

\begin{center}
\textbf{ACKNOWLEDGMENTS}
\end{center}

This work was supported by Ministry of Science and Technology of China (No.
2016YFA0301802); National Natural Science Foundation of China (11504075,
11074062, 11247008, 11374083); Zhejiang Natural Science Foundation
(LZ13A040002); Hangzhou-City Quantum Information and Quantum Optics
Innovation Research Team; The open project funding from CAS Key Laboratory
of Quantum Information, University of Science and Technology of China
(project number KQI201710).

\begin{center}
\textbf{APPENDIX }
\end{center}

Under the unitary operation $U_{1}$ and for an interaction time $t=t_{1}$,
the state transformations for the four logical states $|00\rangle
_{ab},|01\rangle _{ab},|10\rangle _{ab}$ and $|11\rangle _{ab}$ of the two
cat-state qubits are listed below in details.
\begin{align}
& U_{1}|00\rangle _{ab}|g\rangle  \notag \\
& =\exp (i\lambda _{a}\hat{n}_{a}|g\rangle \langle g|t_{1})\exp (i\chi \hat{n%
}_{a}\hat{n}_{b}|g\rangle \langle g|t_{1})\sum\limits_{m=0}^{\infty
}C_{2m}|2m\rangle _{a}\sum\limits_{m^{\prime }=0}^{\infty }C_{2m^{\prime
}}|2m^{\prime }\rangle _{b}|g\rangle  \notag \\
& =\exp (i\lambda _{a}\hat{n}_{a}|g\rangle \langle g|t_{1})\exp (i\chi \hat{n%
}_{a}\hat{n}_{b}|g\rangle \langle g|t_{1})\sum\limits_{m=0}^{\infty
}\sum\limits_{m^{\prime }=0}^{\infty }C_{2m}C_{2m^{\prime }}|2m\rangle
_{a}|2m^{\prime }\rangle _{b}|g\rangle  \notag \\
& =\exp (i\lambda _{a}\hat{n}_{a}t_{1})\exp (i\chi \hat{n}_{a}\hat{n}%
_{b}t_{1})\sum\limits_{m=0}^{\infty }\sum\limits_{m^{\prime }=0}^{\infty
}C_{2m}C_{2m^{\prime }}|2m\rangle _{a}|2m^{\prime }\rangle _{b}|g\rangle
\notag \\
& =\sum\limits_{m=0}^{\infty }\sum\limits_{m^{\prime }=0}^{\infty
}C_{2m}C_{2m^{\prime }}\exp (i\lambda _{a}\hat{n}_{a}t_{1})\exp (i\chi \hat{n%
}_{a}\hat{n}_{b}t_{1})|2m\rangle _{a}|2m^{\prime }\rangle _{b}|g\rangle
\notag \\
& =\sum\limits_{m,m^{\prime }=0}^{\infty }F_{1}\left( m,m^{\prime
},t_{1}\right) C_{2m}C_{2m^{\prime }}|2m\rangle _{a}|2m^{\prime }\rangle
_{b}|g\rangle ,
\end{align}%
\begin{align}
& U_{1}|01\rangle _{ab}|g\rangle  \notag \\
& =\exp \left( i\lambda _{a}\hat{n}_{a}|g\rangle \langle g|t_{1}\right) \exp
(i\chi \hat{n}_{a}\hat{n}_{b}|g\rangle \langle
g|t_{1})\sum\limits_{m=0}^{\infty }C_{2m}|2m\rangle
_{a}\sum\limits_{n^{\prime }=0}^{\infty }C_{2n^{\prime }+1}|2n^{\prime
}+1\rangle _{b}|g\rangle  \notag \\
& =\exp (i\lambda _{a}\hat{n}_{a}|g\rangle \langle g|t_{1})\exp (i\chi \hat{n%
}_{a}\hat{n}_{b}|g\rangle \langle g|t_{1})\sum\limits_{m=0}^{\infty
}\sum\limits_{n^{\prime }=0}^{\infty }C_{2m}C_{2n^{\prime }+1}|2m\rangle
_{a}|2n^{\prime }+1\rangle _{b}|g\rangle  \notag \\
& =\exp (i\lambda _{a}\hat{n}_{a}t_{1})\exp (i\chi \hat{n}_{a}\hat{n}%
_{b}t_{1})\sum\limits_{m=0}^{\infty }\sum\limits_{n^{\prime }=0}^{\infty
}C_{2m}C_{2n^{\prime }+1}|2m\rangle _{a}|2n^{\prime }+1\rangle _{b}|g\rangle
\notag \\
& =\sum\limits_{m=0}^{\infty }\sum\limits_{n^{\prime }=0}^{\infty
}C_{2m}C_{2n^{\prime }+1}\exp (i\lambda _{a}\hat{n}_{a}t_{1})\exp (i\chi
\hat{n}_{a}\hat{n}_{b}t_{1})|2m\rangle _{a}|2n^{\prime }+1\rangle
_{b}|g\rangle  \notag \\
& =\sum\limits_{m,n^{\prime }=0}^{\infty }F_{2}\left( m,n^{\prime
},t_{1}\right) C_{2m}C_{2n^{\prime }+1}|2m\rangle _{a}|2n^{\prime }+1\rangle
_{b}|g\rangle ,
\end{align}%
with%
\begin{eqnarray}
F_{1}\left( m,m^{\prime },t_{1}\right) &=&\exp (i\lambda _{a}2mt_{1})\exp
[i(2m)(2m^{\prime })\chi t_{1}],  \notag \\
F_{2}\left( m,n^{\prime },t_{1}\right) &=&\exp (i\lambda _{a}2mt_{1})\exp
[i(2m)(2n^{\prime }+1)\chi t_{1}].
\end{eqnarray}%
Similarly, one can easily find that
\begin{align}
U_{1}|10\rangle _{ab}|g\rangle & =\exp (i\lambda _{a}\hat{n}_{a}|g\rangle
\langle g|t_{1})\exp (i\chi \hat{n}_{a}\hat{n}_{b}|g\rangle \langle
g|t_{1})\otimes  \notag \\
& \;\;\,\sum\limits_{n=0}^{\infty }C_{2n+1}|2n+1\rangle
_{a}\sum\limits_{m^{\prime }=0}^{\infty }C_{2m^{\prime }}|2m^{\prime
}\rangle _{b}|g\rangle  \notag \\
& =\sum\limits_{n,m^{\prime }=0}^{\infty }F_{3}\left( n,m^{\prime
},t_{1}\right) C_{2n+1}C_{2m^{\prime }}|2n+1\rangle _{a}|2m^{\prime }\rangle
_{b}|g\rangle ,
\end{align}%
\begin{align}
U_{1}|11\rangle _{ab}|g\rangle & =\exp (i\lambda _{a}\hat{n}_{a}|g\rangle
\langle g|t_{1})\exp (i\chi \hat{n}_{a}\hat{n}_{b}|g\rangle \langle
g|t_{1})\otimes  \notag \\
& \;\;\,\sum\limits_{n}^{\infty }C_{2n+1}|2n+1\rangle _{a}\otimes
\sum\limits_{n^{\prime }=0}^{\infty }C_{2n^{\prime }+1}|2n^{\prime
}+1\rangle _{b}|g\rangle  \notag \\
& =\sum\limits_{n,n^{\prime }=0}^{\infty }F_{4}\left( n,n^{\prime
},t_{1}\right) C_{2n+1}C_{2n^{\prime }+1}|2n+1\rangle _{a}|2n^{\prime
}+1\rangle _{b}|g\rangle ,
\end{align}%
with
\begin{eqnarray}
F_{3}\left( n,m^{\prime },t_{1}\right) &=&\exp [i\lambda
_{a}(2n+1)t_{1}]\exp [i(2n+1)(2m^{\prime })\chi t_{1}],  \notag \\
F_{4}\left( n,n^{\prime },t_{1}\right) &=&\exp [i\lambda
_{a}(2n+1)t_{1}]\exp [i(2n+1)(2n^{\prime }+1)\chi t_{1}].
\end{eqnarray}


\begin{thebibliography}{99}
\bibitem{58} For a transmon qutrit, the $|g\rangle$ $\leftrightarrow$ $%
|f\rangle$ transition is much weaker than those of the $|g\rangle$ $%
\leftrightarrow$ $|e\rangle$ and $|e\rangle$ $\leftrightarrow$ $|f\rangle$
transitions. Thus, we have $\gamma^{-1}_{fg} \gg \gamma^{-1}_{eg},
\gamma^{-1}_{fe}$.
\end{thebibliography}
\end{document}